\documentclass[lettersize,journal]{IEEEtran}
\usepackage{amsmath,amsfonts}
\usepackage{algorithmic}
\usepackage{algorithm}
\usepackage{array}
\usepackage{textcomp}
\usepackage{stfloats}
\usepackage{url}
\usepackage{verbatim}
\usepackage{graphicx}
\usepackage{cite}
\usepackage{multirow}
\usepackage{subcaption}
\usepackage{float}  
\usepackage{caption}
\usepackage{booktabs}
\usepackage{diagbox}
\usepackage{makecell}
\usepackage{color}
\usepackage{fancyhdr}
\allowdisplaybreaks[4]

\begin{document}

\title{Deep Learning Assisted Multiuser MIMO Load Modulated Systems for Enhanced Downlink mmWave Communications}

\author{Ercong Yu, Jinle Zhu, Qiang Li,
Zilong Liu,~\IEEEmembership{Senior Member,~IEEE,}
Hongyang Chen,~\IEEEmembership{Senior Member,~IEEE,}
Shlomo Shamai (Shitz),~\IEEEmembership{Life Fellow,~IEEE,}
and H. Vincent Poor,~\IEEEmembership{Life Fellow,~IEEE}
\thanks{E. Yu, J. Zhu, and Q. Li are with the Yangtze Delta Region Institute of University of Electronic Science and Technology of China, Huzhou 313098, China, and also with the National Key Laboratory of Science and Technology on Communications, University of Electronic Science and Technology of China (UESTC), Chengdu 611731, China, e-mails: ercong-kang@outlook.com; sophia\_zhujl@163.com; liqiang@uestc.edu.cn.
(\emph{Corresponding author: Qiang Li.})}
\thanks{Z. Liu is with the School of Computer Science and Electronics Engineering, University of Essex, UK, e-mail: zilong.liu@essex.ac.uk.}
\thanks{H. Chen is with the Research Center for Graph Computing, Zhejiang Lab, Hangzhou 311100, China, email: dr.h.chen@ieee.org; hongyang@zhejianglab.com.}
\thanks{Shlomo Shamai (Shitz) is with the Technion-Israel Institute of Technology, Haifa 320003, Israel e-mail: sshlomo@ee.technion.ac.il.}
\thanks{H. V. Poor is with the Department of Electrical Engineering, Princeton University, Princeton, NJ, 08544, USA, e-mail: poor@princeton.edu.}}



\maketitle

\thispagestyle{fancy}
\fancyhead{}
\lhead{}
\rhead{}
\lfoot{}
\cfoot{}   
\rfoot{}
\fancyhead[L]{\small \copyright 20xx IEEE. Personal use of this material is permitted. Permission from IEEE must be obtained for all other uses, in any current or future media, including reprinting/republishing this material for advertising or promotional purposes, creating new collective works, for resale or redistribution to servers or lists, or reuse of any copyrighted component of this work in other works.}
\renewcommand{\headrulewidth}{0pt}

\begin{abstract}

  This paper is focused on multiuser load modulation arrays (MU-LMAs) which are attractive due to their low system complexity and reduced cost for millimeter wave (mmWave) multi-input multi-output (MIMO) systems.
  The existing precoding algorithm for downlink MU-LMA relies on a sub-array structured (SAS) transmitter which may suffer from decreased degrees of freedom and complex system configuration.
  Furthermore, a conventional LMA codebook with codewords uniformly distributed on a hypersphere may not be channel-adaptive and may lead to increased signal detection complexity. 
  In this paper, we conceive an MU-LMA system employing a full-array structured (FAS) transmitter and propose two algorithms accordingly.
  The proposed FAS-based system addresses the SAS structural problems and can support larger numbers of users.
  For LMA-imposed constant-power downlink precoding, we propose an FAS-based normalized block diagonalization (FAS-NBD) algorithm.
  However, the forced normalization may result in performance degradation.
  This degradation, together with the aforementioned codebook design problems, is difficult to solve analytically. This motivates us to propose a Deep Learning-enhanced (FAS-DL-NBD) algorithm for adaptive codebook design and codebook-independent decoding.
  It is shown that the proposed algorithms are robust to imperfect knowledge of channel state information and yield excellent error performance. Moreover, the FAS-DL-NBD algorithm enables signal detection with low complexity as the number of bits per codeword increases.

\end{abstract}


\section{Introduction} \label{S_intro}

\IEEEPARstart{T}{he} millimeter-wave (mmWave) bands hold a promising prospect for next-generation wireless communications due to the abundant bandwidth and the potential to offer high data rates. 
The small wavelengths at mmWave bands permit the use of a massive antenna array in a collocated area as well as multiple antenna technologies such as multiple-input multiple-output (MIMO) \cite{book1}. 
Of course, MIMO systems have attracted significant attention due to their diversity and multiplexing gains \cite{ref_n1, D1, MIMO}. 
However, the use of large numbers of antennas in conventional MIMO systems can result in prohibitively high system complexity and hardware cost as each transmit antenna requires a separate radio frequency (RF) chain and an associated power amplifier (PA). 
In a practical MIMO system, these PAs distributed on each transmit antenna impose per-antenna power constraints \cite{ref_n7,n3,n4}. Despite the fact that convex optimization methods and the capacity region duality could be adapted to downlink channels with per-antenna power constraints \cite{n4}, the relevant optimization is still challenging \cite{n3}. 
Moreover, voltage modulation of a conventional MIMO system may impose a linearity requirement on the PAs for improved power efficiency. 
An effective solution to circumvent these drawbacks is to develop a communication system based on load modulation arrays (LMAs) \cite{D2, D3,ref12}. 

Unlike the conventional MIMO transmitter, an LMA transmitter uses a central power amplifier (CPA) to serve the entire antenna array with any number of antennas. 
By feeding the CPA using a single source with a fixed voltage level and frequency, the transmitted signal is modulated via varying the antenna load impedance in accordance with information bits directly \cite{D3}. 
In this way, the LMA transmitter eliminates the need for an RF chain per antenna and thus avoids the problems of per-antenna power constraints.
As the number of antennas in massive MIMO systems grows, the use of an LMA transmitter leads to a significant reduction in the RF chain cost and system complexity accordingly.
However, the mismatch in antenna impedances may cause power flow back to the CPA which could decrease the power efficiency.
To address this issue, it is desirable that the instantaneous sum power at the transmitter should be constant \cite{D6}. 

\subsection{Related Works}

From the precoding perspective, most existing algorithms target sum power constraints \cite{n11, n22, n33} or per-antenna power constraints \cite{n3, ref_n7}. However, unlike the aforementioned two types of constraints where the capacity region is known \cite{n2}, the capacity region of the downlink channels with constant power constraints has not been well studied, and little is known about the relevant precoding algorithms. 
For a downlink multiuser LMA (MU-LMA) communication system with constant power constraint, \cite{LSE} proposed an iterative precoding algorithm based on the framework of least square error. However, it is only valid for systems where each user equipment (UE) has a single receive antenna. 
By relaxing the constraint on the number of receive antennas, \cite{ref_n8} developed a precoding algorithm based on a sub-array structured (SAS) transmitter. Such an algorithm can ensure the power constraint by configuring an exclusive LMA transmit unit for each user and then eliminating the multiuser interference (MUI) using the block diagonalization (BD) algorithm proposed in \cite{BD}. 

\begin{figure*}[!htpb]
  
  \centering
  \includegraphics[scale=0.95]{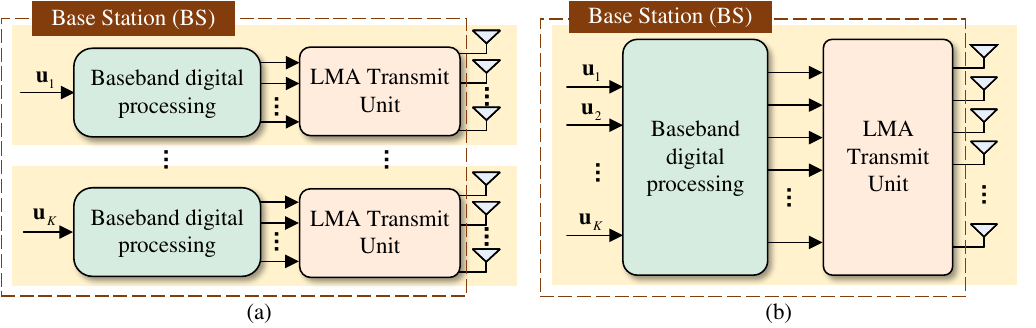}
  \caption{Structures of two types of MU-LMA transmitters. (a) SAS transmitter; (b) FAS transmitter.} \label{cp}
  \end{figure*}

However, an MU-LMA system employing an SAS transmitter may not be able to support a large number of users and suffers from the following \emph{structure-related} problems. To understand this, let us look at the SAS transmitter shown in Fig.~\ref{cp}a. First, as each user is assigned a part of the antenna array, the system's degrees of freedom shrink. 
This results in a deteriorated bit error rate (BER) performance when the number of users increases \cite{book2}. 
Second, as the precoding matrices between users are forced to be diagonally arranged, the total number of transmit antennas must be an integer multiple of the number of users. This imposes inflexibility in system configuration. 
Moreover, the minimum number of transmit antennas required to support a given number of users grows quadratically, which severely limits the number of concurrent users supported by the system. 
In addition, the design of the combiner is absent in the SAS-precoding, which limits the algorithm's effectiveness for varying numbers of receive antennas. 

In communication systems employing LMAs, codebook design is another crucial issue. 
Phase Modulation on the Hypersphere (PMH) is regarded as a generalized form of signal codebook generation for LMA systems. It ensures a constant instantaneous sum power by designing a set of points distributed on the surface of a multidimensional hypersphere via clustering methods \cite{KMeans, ref_n2}.
It is known that the capacity of PMH on the additive white Gaussian noise channel is achieved with points distributed uniformly on the surface of a multidimensional hypersphere \cite{multi}. 
However, so far, the LMA codebook design for mmWave channels is largely open, and the uniformly distributed PMH method is not optimal for downlink mmWave channels. In particular: 1) the generated codebook points may be inflexible, unable to adapt to fading channels; 2)  as the number of transmitted bits increases, the size of the codebook increases exponentially, resulting in an exponential increase of complexity for codebook generation and signal detection. 

That said, the \emph{codebook-related} problem (i.e., codebook design for mmWave channels, complexity increase in terms of codebook design and detection) may not be regarded as a simple optimization problem. Therefore, we employ deep learning (DL), which is considered a powerful tool that mitigates challenges in MIMO communication systems \cite{LR_DL}, and tackle this problem from a new perspective.
The end-to-end (E2E) learning concept was first proposed in \cite{ref_n9}. As a holistic approach to designing the transmitter and receiver in one step, the end-to-end learning system seeks to find the optimal solution for the entire system, as opposed to the optimal solution for each separate block. Supervised by the objective of the recurrence of transmitted signals, the trained encoder in an end-to-end system is capable of generating codebooks adapting to given channels, thus improving the bit-level precision \cite{ref_n13}. 
Additionally, a trained decoder operates the signal detection independent of the codebook size and thus simplifies the detection complexity when transmitting a large number of information bits at a time. 

On the other hand, an end-to-end network can theoretically be used to construct the entire downlink MU-LMA system, thereby achieving the MUI cancellation and the codebook-related problem in one step. However, this would lead to convergence difficulty due to the conflict of multiple tasks. In this paper, we advocate the idea of constructing an end-to-end network that focuses only on the codebook-related problem. 

\subsection{Contributions}

In view of the above background, this paper develops an enhanced transmitter structure, new codebooks, and a signal detection method with reduced complexity for MU-LMA systems for downlink mmWave channels. 

First, we propose a new MU-LMA communication system employing a full-array structured (FAS) transmitter. 
As shown in Fig.~\ref{cp}b, unlike the SAS transmitter, all the users in the FAS transmitter share the entire antenna array. Thus, the degrees of freedom per user (which are independent of the number of users) increase. Further, the FAS transmitter can dynamically support varying numbers of users without imposing a proportional relationship between the number of transmit antennas and the number of users. Furthermore, it can break the upper limit of the number of users supported by an SAS transmitter \cite{ref_n8}. 

Subsequently, we consider an FAS-based normalized BD (FAS-NBD) algorithm. We address the MUI cancellation problem using the well-known BD proposed in \cite{BD} and then design a normalization module to achieve the constant power constraint. The FAS-NBD algorithm is LMA-adaptive and can jointly design the precoders and combiners. However, the forced normalization may cause BER \emph{performance degradation} \cite{ref_n8}. 

To alleviate the performance degradation and address the codebook-related problem, we develop a novel FAS-based DL-enhanced normalized BD (FAS-DL-NBD) algorithm. 
Instead of using the conventional codebook (i.e., a set of uniformly distributed PMH points) at the transmitter, we deploy a multilayer fully connected feedforward neural network (FC-FNN) as an \emph{encoder} before the BD precoder. Such encoders seek to generate codebooks adapting to the fading channels and alleviating the performance degradation in FAS-NBD. 
Likewise, we deploy a multilayer FC-FNN as a \emph{decoder} at each receiver to replace the conventional maximum likelihood (ML) detector. Since a trained decoder is independent of the codebook size, it leads to improved LMA signal detection with low complexity. 
As multiple FC-FNNs are nested on different parts of the network, the framework of the proposed FAS-DL-NBD algorithm can be regarded as an E2E-like FC-FNN-reinforced communication network. In contrast to the conventional one-step end-to-end network \cite{ref_n9}, the nested FC-FNNs cooperate with the NBD precoder and are trained with refined objectives. 
This ensures the convergence of the network and reduces the difficulty of training. 

\begin{table*}[!htpb]
  \setlength{\abovecaptionskip}{0pt}
  \setlength{\belowcaptionskip}{0pt}
  \centering
  \caption{List of acronyms.}
  \label{T_A}
  \begin{tabular}{cccc}
  \toprule
  \textbf{Acronyms} & \textbf{Description} & \textbf{Acronyms} & \textbf{Description} \\  
  \midrule
  BD	&	block diagonalization	&	LMA	&	load modulation array	\\
  BER	&	bit error rate	&	MIMO	&	multi-input multi-output	\\
  BS	&	base station	&	ML	&	maximum likelihood	\\
  CPA	&	central power amplifier	&	mmWave	&	millimeter wave	\\
  CSI	&	channel state information	&	MU	&	multiuser	\\
  DL	&	deep learning	&	MUI	&	multiuser interference	\\
  FAS	&	full-array structured	&	PASPR	&	peak-to-average sum power ratio	\\
  FAS-NBD	&	FAS-based normalized BD	&	PMH	&	phase modulation on the hypersphere	\\
  FAS-DL-NBD	&	FAS-based DL-enhanced normalized BD	&	RF	&	radio frequency	\\
  FAS-E2E	&	FAS-based E2E learning	&	SAS	&	sub-array structured	\\
  FC-FNN	&	fully connected feedforward neural network	&	SNR	&	signal-to-noise ratio	\\
  ICSI	&	imperfect knowledge of CSI	&	SVD	&	singular value decomposition	\\
  LM & load modulator	&	UE	&	user equipment	\\
  \bottomrule
  \end{tabular}
  \end{table*}
  
The superiority of the proposed FAS-based system is proven with theoretical analysis. Meanwhile, we compare the performance of the proposed FAS-DL-NBD algorithm and the one-step end-to-end learning in terms of convergence capability. The performance of the two proposed algorithms is compared with that of the existing SAS-precoding algorithm in terms of the bit error rate (BER) and the robustness against imperfect knowledge of channel state information (ICSI). We also show the advantages of the two proposed algorithms to support a larger number of users. Further, we demonstrate the capability of the DL-enhanced algorithm with achieving a low-complexity detection by varying the number of transmitted information bits.
  
\subsection{Organizations and Notation}

The remainder of this paper is organized as follows. The system model is explained in Section~\ref{S_system}. The frameworks of FAS-NBD and FAS-DL-NBD are illustrated in Sections~\ref{S_BD} and \ref{S_DL}, respectively. 
Section~\ref{S_disc1} discusses the advantages of the proposed FAS-based algorithms compared with the SAS-precoding. Section~\ref{S_disc2} examines the convergence of the FAS-DL-NBD and its advantages over an LMA-adaptive E2E-based framework.
Section~\ref{S_sim} presents our simulation results. Finally, we conclude this paper in Section~\ref{S_con} by summarizing the performance of the proposed algorithms and presenting our conclusions.

\emph{Notation:} Scalars are represented by italicized characters, while matrices and vectors are represented by bold upper case and lower case characters. Uppercase calligraphic letters represent specially defined sets, such as $\mathcal{S}$. Matrix elements are represented by $[\cdot]$, whereas set elements are represented by $\{\cdot\}$. The Frobenius norm of matrices or vectors is represented by $\parallel\cdot\parallel$. Furthermore, the set of real and complex-valued numbers are denoted respectively by the symbols $\mathbb{R}$ and $\mathbb{C}$. $\mathbf{A}\in\mathbb{C}^{M\times N}$ ($\mathbf{A}\in\mathbb{R}^{M\times N}$) denotes that $\mathbf{A}$ is a complex-valued (real-valued) matrix with $M$ rows and $N$ columns. A complex Gaussian random variable is denoted by $\mathcal{CN}\sim(\mu, \sigma^2)$, where $\mu$ is the mean and $\sigma^2$ is the variance. Transpose and Hermitian transpose are also represented by $(\cdot)^T$ and $(\cdot)^H$, respectively. Furthermore, $\left\lceil \cdot \right\rceil$ and $\left\lfloor \cdot \right\rfloor$ represent the ceiling and floor functions, respectively.

For the convenience of readers, the acronyms used in this paper are listed in Table~\ref{T_A}.

\section{Multiuser LMA Model on the Downlink} \label{S_system}

\subsection{Load Modulated Arrays} \label{S_LMA}

\begin{figure}[!htpb]
  
  \centering
  \includegraphics[scale=0.95]{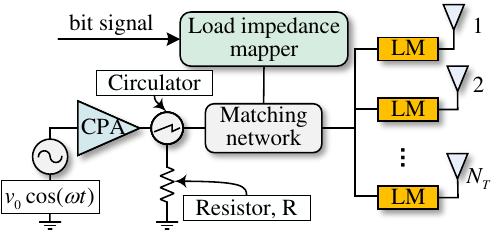}
  \caption{Structure of the Load Modulated MIMO Transmitter.} \label{LMA}
  \end{figure}

The structure of an LMA MIMO transmitter with $N_T$ antennas is depicted in Fig.~\ref{LMA}.  
The CPA serves the entire antenna array and is powered by a constant-magnitude RF carrier source. Each antenna is equipped with an LM, which can be implemented with varactor diodes or pin diodes. 
Assume the impedance on the $i$th antenna is $Z_i$. The $N_T\times 1$ load impedance vector could be represented as $\mathbf{Z}=[Z_1,\cdots,Z_{N_T}]^T$. 
As the voltage magnitude is always constant, the current on the $i$th antenna is proportional to $1/Z_i$. 
By selecting an impedance vector in accordance with given information-bearing bits, the antenna currents vary and thus result in a modulated transmit signal. Consequently, an LMA transmitter utilizing pin diodes necessitates only a level shifter to connect the digital baseband to the pin-diode switches, as opposed to the DACs, upconverters, and mixers required by conventional transmitter structures.

At the transmitter, the effective admittance seen by the power source is $Y=\sum^{N_T}_{i=1}\frac{1}{Z_i}$. Notably, the LMA may suffer a severe mismatch between the varying load impedances of different antennas and the effective antenna load impedance. This leads to reduced energy efficiency as power could flow back to the CPA. As shown in Fig.~\ref{LMA}, to redirect any reflected power to the resistor R, a circulator is employed. The CPA efficiency is described by the PASPR \cite{D6}. It is the peak-to-average power ratio aggregated over all the antenna elements. 
To address the reduced energy efficiency issue and ensure the PASPR of 1, the LMA signal vectors should be distributed on the surface of a multidimensional hypersphere (i.e., PMH) \cite{ref_n3}.

For an LMA communication system, the PMH codebook $\mathcal{S}_{n, P}$ with constant power constraint can be expressed as
\begin{equation}
  \mathcal{S}_{n, P}=\{\mathbf{t}_i\in \mathbb{C}^{n\times 1}|\parallel\mathbf{t}_i\parallel^2=P\}, \label{e_PMH1}
\end{equation}
where the number of information bits is denoted as $n$, and the power for transmission is constrained to $P$. $M=2^n$ stands for the codebook size, and $\mathbf{t}_i\in \mathbb{C}^{n\times 1}$ denotes the $i$th codeword. The construction of a conventional LMA codebook $\mathcal{S}_{n, P}$ can be formulated as a spherical code construction problem \cite{D6}:
\begin{equation}
  \max_{\mathcal{S}_{n, P}\subset \mathcal{C}_{n, P}}\left(\min_{\mathbf{t}_i,\mathbf{t}_j\in \mathcal{S}, i\neq j} \parallel\mathbf{t}_i-\mathbf{t}_j\parallel\right), \label{e_PMH2}
\end{equation}
where $\mathcal{C}_{n, P}$ is the set of points distributed on the surface of a multidimensional hypersphere with a dimension of $n$ and a radius of $\sqrt{P}$. $\mathcal{S}_{n, P}$ is a subset of $\mathcal{C}_{n, P}$ where the minimum distance between each pair of the element points is maximized.

\subsection{Proposed Multiuser LMA MIMO System} \label{sys_model}

\begin{figure*}[!t]
  
  \centering
  \includegraphics[scale=0.95]{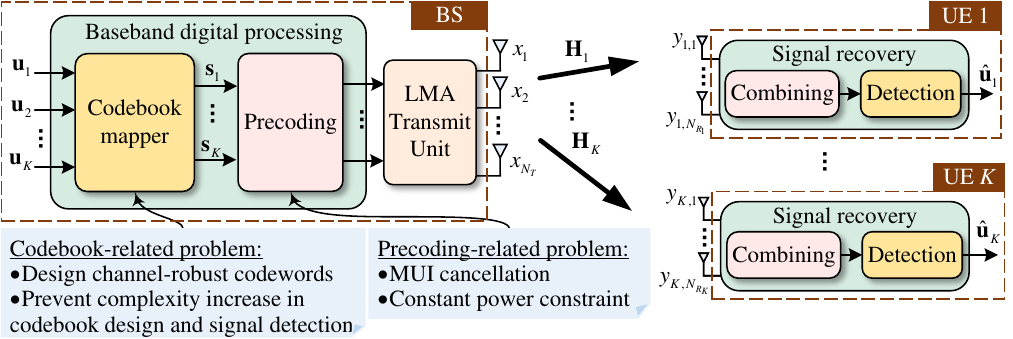}
  \caption{Structure of the Proposed FAS-based MU-LMA System\protect\footnotemark.}
  \label{fig_sys}
  \end{figure*}

We conceive a downlink MU-LMA communication system employing an FAS transmitter. As shown in Fig.~\ref{fig_sys}, the BS transmits signals to $K$ users through $N_T$ antennas. At the receiver, the $k$th user, denoted as $U_k~(k=1,\cdots, K)$, is equipped with $N_{R_k}$ antennas, and the total number of receive antennas is denoted as $N_R=\sum_{k=1}^KN_{R_k}$.
Assume $U_k$ has $n_k$ bits to send where the corresponding information bit vector is represented as $\mathbf{u}_k=[u_{k,1},\cdots, u_{k,n_k}]^T$. The total number of bits for all users is represented as $N=\sum^K_{k=1}n_k$. 

First, the information bits for $U_k$ are encoded into a complex-valued vector $\mathbf{s}_k$ in accordance with a given codebook. Then, the composite coded vector for all users can be represented as $\mathbf{s}=[\mathbf{s}_1^T,\cdots,\mathbf{s}_K^T]^T\in \mathbb{C}^{N\times 1}$.
In view of the design of an LMA codebook, a basic method is to generate signal vectors distributed uniformly on the surface of a hypersphere. However, this method may lead to increased ML detection complexity and poor adaptability to fading channels. To address this issue and generate robust codewords, the design of the codebook can be optimized which is called the \emph{codebook-related} problem.
Next, regarding the multiuser downlink scenario, a precoding algorithm is required to achieve MUI cancellation and constant power constraints (i.e., $\mathbf{x}^H\mathbf{x}=P$), i.e., the \emph{precoding-related} problem.
The precoding and codebook-related problems will be addressed in Section~\ref{S_BD} and Section~\ref{S_DL}, respectively.

\footnotetext{The modules at the transmitter cooperate with the corresponding modules at the receiver to achieve the specified goals, but for the sake of brevity and clarity, the wiring of the receiver module has been simplified in this diagram.}

After the encoding and precoding processes, the signal to be transmitted is denoted as $\mathbf{x}=[x_1,\cdots,x_{N_T}]^T$. Assume the channel matrix of $U_k$ is represented as $\mathbf{H}_k$. Then, the received signal $\mathbf{y}_k=[y_{k,1},\cdots,y_{k,N_{R_k}}]^T$ at $U_k$ is represented as
\begin{equation}
  \mathbf{y}_k=\mathbf{H}_k\mathbf{x}+\mathbf{n}_k,
\end{equation}
where $\mathbf{n}_k\sim \mathcal{CN}(0,\sigma^2)$ stands for the noise at $U_k$. 
Finally, after the combining and detection process, the information bits recovered at the receiver of $U_k$ is denoted as $\hat{\mathbf{u}}_k=[\hat{u}_{k,1},\cdots, \hat{u}_{k,n_k}]^T$.

\subsection{Channel Model}

We consider a multipath but no clustered narrowband mmWave channel between the BS and the UE.
Assume the number of scattering paths is $N_{ray}$. According to the system model defined in Subsection~\ref{sys_model}, the normalized narrowband mmWave channel of $U_k$ is modelled as
\begin{equation}
\mathbf{H}_k = \sqrt{\frac{N_TN_{R_k}}{N_{ray}}}\sum\limits_{l=1}^{N_{ray}}\alpha_k^l
\mathbf{a}_r\left(\theta_k^l\right)\mathbf{a}_t^\dagger\left(\phi_k^l\right),
\end{equation}
where $\alpha_k^l\sim \mathcal{CN}(0,1) $ is the channel gain of the $l$th path of $U_k$. Meanwhile, $\theta_k^l \text{ and } \phi_k^l$ represent the angle of departure and angle of arrival, respectively. 
Further, uniform linear arrays are employed to represent the response vector at UE and BS, which are denoted as $\mathbf{a}_r\left(\theta\right)$ and $\mathbf{a}_t\left(\phi\right)$), respectively, i.e., 
\begin{equation}
\mathbf{a}_r\left(\theta\right) = \frac{1}{{\sqrt {N_{R_k}} }}{\left[ {1,{e^{j\frac{{2\pi {\rm{d}}}}{\lambda }\cos \left( \theta  \right)}},...,{e^{j\frac{{\left( {{N_{R_k}} - 1} \right)2\pi {\rm{d}}}}{\lambda }\cos \left( \theta  \right)}}} \right]^T},
\end{equation}
\begin{equation}
\mathbf{a}_t\left(\phi\right) = \frac{1}{{\sqrt {N_{t}} }}{\left[ {1,{e^{j\frac{{2\pi {\rm{d}}}}{\lambda }\cos \left( \phi  \right)}},...,{e^{j\frac{{\left( {{N_{t}} - 1} \right)2\pi {\rm{d}}}}{\lambda }\cos \left( \phi  \right)}}} \right]^T},
\end{equation}
where $\lambda$ is the carrier wavelength, and $d$ denotes the antenna spacing.
Note that the system structure and algorithms proposed in this paper can be generalized to other channel models.

\section{Proposed Precoding Algorithm Based on BD} \label{S_BD}

Aiming for addressing the precoding-related problem shown in Fig.~\ref{fig_sys}, we propose an LMA-adaptive precoding algorithm in this section. The proposed FAS-NBD algorithm achieves MUI cancellation using BD and addresses the constant power constraint with normalization in turn. Precoders and combiners are jointly designed using this algorithm. 

\subsection{MUI Cancellation}
We design a precoding matrix at the transmitter to eliminate the MUI and thus maximize the system capacity. The precoding matrix $\mathbf{F}$ for $K$ users is formulated as
\begin{equation}
  \mathbf{F}=[\mathbf{F}_1, \mathbf{F}_2, \cdots, \mathbf{F}_K]\in \mathbb{C}^{N_T\times N},
\end{equation}
where $\mathbf{F}_k\in \mathbb{C}^{N_T\times n_k}$ is the precoding matrix for $U_k$. 

In this algorithm, the coded vector $\mathbf{s}_k$ for $U_k$ is selected from a PMH codebook $\mathcal{S}_{n_k, P}$ (as shown in~(\ref{e_PMH1}) and (\ref{e_PMH2})). The codewords are distributed uniformly on the surface of a hypersphere, which can be achieved using K-means clustering \cite{KMeans}.
Assume the composite coded signal for all users is denoted as $\mathbf{s}=[\mathbf{s}_1^T,\cdots,\mathbf{s}_K^T]^T$. 
Then, the received signal for $U_k$ is represented as 
\begin{align}
  \mathbf{y}_k =\mathbf{H}_k\mathbf{F}\mathbf{s}+\mathbf{n}_k \nonumber = \mathbf{H}_k\mathbf{F}_k\mathbf{s}_k+\mathbf{H}_k\sum\limits_{i\neq k}^K\mathbf{F}_i\mathbf{s}_i+\mathbf{n}_k, \label{receiveSig}
  \end{align}
where $\sum\limits_{i\neq k}^K\mathbf{F}_i\mathbf{s}_i$ denotes the MUI of $U_k$ for the downlink communication which should be minimized. For $1\leq k\neq i \leq K$, The \emph{MUI cancellation} problem is formulated as 
\begin{equation}
\mathbf{H}_k\mathbf{F}_i =
\left\{
\begin{aligned}
&\mathbf{0}, &{~~k\neq i} \\
&\mathbf{H}_k\mathbf{F}_k , &{~~k=i}
\end{aligned}~.
\right. \label{P1}
\end{equation}
Here, $\mathbf{F}$ is said to block diagonolize $\mathbf{H}$, where $\mathbf{H}=[\mathbf{H}_1^T, \mathbf{H}_2^T, \cdots, \mathbf{H}_K^T]^T \in \mathbb{C}^{N_R\times N_T}$. 
This indicates that the precoder of the $k$th user should be in the null space of other user channels. We define the composite channel of users except $U_k$ as $\widetilde{\mathbf{H}}_k=[\mathbf{H}_1^T, \cdots, \mathbf{H}_{k-1}^T,$ $\mathbf{H}_{k+1}^T, \cdots, \mathbf{H}_K^T]^T\in \mathbb{C}^{(N_R-N_{R_k})\times N_T}$.
$\mathbf{F}_k$ should lie in the null space of $\widetilde{\mathbf{H}}_k$.
The SVD of $\widetilde{\mathbf{H}}_k$ is given by
\begin{equation}
\widetilde{\mathbf{H}}_k = \widetilde{\mathbf{U}}_k\widetilde{{\bf\Sigma}}_k\left[\widetilde{\mathbf{V}}_k^1 ~~ \widetilde{\mathbf{V}}_k^0\right]^H,
\end{equation}
where $\widetilde{\mathbf{V}}_k^0$ consists of the last $N_T-(N_R-N_{R_k})$ columns of the right singular vectors and is the basis of the null space of $\widetilde{\mathbf{H}}_k$. The existence of $\widetilde{\mathbf{V}}_k^0$ is ensured by
\begin{equation}
  N_T-(N_R-N_{R_k}) > 0. \label{Dim1}
\end{equation}

With the MUI canceled by $\widetilde{\mathbf{V}}_k^0$, the users' channels can be separated as independent channels.
Given the compact channel of all users, this can be presented as
\begin{align}
\mathbf{H}\left[\widetilde{\mathbf{V}}_1^0, \widetilde{\mathbf{V}}_2^0, \cdots,\widetilde{\mathbf{V}}_K^0\right]&=
\begin{bmatrix}
\mathbf{H}_1\\
\vdots \\
\mathbf{H}_K
\end{bmatrix}
\left[\widetilde{\mathbf{V}}_1^0, \widetilde{\mathbf{V}}_2^0, \cdots,\widetilde{\mathbf{V}}_K^0\right]
\\&=\begin{bmatrix}
  \mathbf{H}_1\widetilde{\mathbf{V}}_1^0 & \mathbf{0} & \cdots & \mathbf{0} \\ 
  \mathbf{0} & \mathbf{H}_2\widetilde{\mathbf{V}}_2^0 & \cdots & \mathbf{0} \\  
  \vdots & \vdots & \ddots & \vdots \\  
  \mathbf{0} & \mathbf{0} & \cdots & \mathbf{H}_K\widetilde{\mathbf{V}}_K^0
\end{bmatrix}.
\end{align}
Consequently, the subsequent precoding and combining matrix can be derived from the SVD of their equivalent channels, which is given as 
 \begin{equation}
\mathbf{H}_k{\widetilde{\mathbf{V}}}_k^0 = \left[\bar{\mathbf{U}}_k^1~~\bar{\mathbf{U}}_k^0\right]\bar{{\bf\Sigma}}_k^t\left[\bar{{\mathbf{V}}}_k^1~~\bar{{\mathbf{V}}}_k^0\right]^H, \label{Eq_span}
 \end{equation}
where $\bar{{\mathbf{V}}}_k^1$  and $\bar{\mathbf{U}}_k^1$ are formed by the first $n_k$ columns of the right singular matrix and the left singular matrix, respectively.
$\bar{{\mathbf{V}}}_k^1$ is the basis of the equivalent channel $\mathbf{H}_k{\widetilde{\mathbf{V}}}_k^0$. 
It presents the directions where the signal of $U_k$ has the most span and is used to enhance the signal towards the corresponding channel. Then, similar to (\ref{Dim1}), the existence of $\bar{{\mathbf{V}}}_k^1$ and $\bar{\mathbf{U}}_k^1$ is guaranteed by
\begin{align}
  n_k\leq N_{R_k} \leq N_T-N_R+N_{R_k}.
\end{align}
As a result, the sufficient condition for the existence of the FAS-NBD precoding matrix is 
\begin{align}
  \begin{cases}
    N_T\geq  N_R \\  
    n_k\leq N_{R_k}.
  \end{cases}
\end{align}

Therefore, the \emph{BD precoder} can be given by
\begin{align}
\mathbf{F} &= \left[{\widetilde{\mathbf{V}}}_1^0,\cdots,{\widetilde{\mathbf{V}}}_K^0\right]
\begin{bmatrix}
\bar{{\mathbf{V}}}_1^1 & \cdots & \mathbf{0}\\
\vdots & \ddots & \vdots \\
\mathbf{0} & \cdots & \bar{{\mathbf{V}}}_K^1
\end{bmatrix}
\\&=\begin{bmatrix}
{\widetilde{\mathbf{V}}}_1^0\bar{{\mathbf{V}}}_1^1,\cdots,{\widetilde{\mathbf{V}}}_K^0\bar{{\mathbf{V}}}_K^1
\end{bmatrix}. \label{BD_matrix}
\end{align}
Sequentially, the \emph{BD combiner} for $U_k$ is $\mathbf{W}_k = (\bar{\mathbf{U}}_k^1)^H$.

\subsection{Normalization}

Notably, although the precoding matrix $\mathbf{F}_k$ of each user is unitary, their composite matrix $\mathbf{F}$ is not unitary and therefore not norm-preserving. Therefore, the precoded signal $\mathbf{Fs}$ may result in varying sum power within a small range. Assume the power for transmission is constrained to a given power $P_T$. Then the transmitted signal vector $\mathbf{x}$ is normalized as 
\begin{equation}
  \mathbf{x} = \frac{\sqrt{P_T}}{\parallel \mathbf{Fs}\parallel}\mathbf{Fs}.
\end{equation}
$\frac{\sqrt{P_T}}{\parallel \mathbf{Fs}\parallel}$ is called the \emph{normalization factor}. Such a factor depends on the combination of all users' transmission signals and fluctuates within a very narrow range of approximately 1. Normalization is essential for ensuring the power efficiency of the LMA transmitter.

\subsection{Signal Detection}
Based on the ML criterion, the \emph{signal detection} at $U_k$ is presented as
\begin{align}
&\mathbf{s}_k^\star = \min \|\mathbf{W}_k\mathbf{H}_k\mathbf{F}_k\mathbf{s}_t-\mathbf{W}_k\mathbf{y}_k\|\\
&s.t.~\mathbf{s}_t\in{\mathcal{S}_{n_k, P}},
\end{align}
where $\mathbf{W}_k\mathbf{y}_k$ is the received signal after combining operation. $\mathbf{s}_k^\star$ is a signal vector detected with reference to the codebook of $U_k$. Finally, the bit information can be obtained.

\subsection{Algorithm Limitations}

The normalization factor, which determines the power scaling of transmitted signals for each user, is floating and agnostic for the UE side. Its value depends on the real-time combination of all user signals and may cause a slight degradation in performance.
This is a trade-off in terms of system flexibility and degree-of-freedom gains. 
In fact, as the number of users in the MU-LMA system grows, the increased degree-of-freedom gain and the system flexibility could outweigh the system performance.
Furthermore, this performance degradation will be optimized in Section~\ref{S_DL} as an additional issue for the codebook-related problem. 

In addition, due to the relationship between codebook size and the number of information bits (i.e., $M=2^n$), the signal detection and codebook design complexity of the FAS-NBD algorithm may increase exponentially when transmitting numerous bits. This will also be addressed in Section~\ref{S_DL} with a trained network. 

\section{Proposed DL-enhanced Algorithm} \label{S_DL}

In this section, the FAS-DL-NBD algorithm is proposed to further address the performance degradation and codebook-related problem shown in Fig.~\ref{fig_sys}. Instead of the conventional codebook and ML detection used in the FAS-NBD algorithm, multilayer FC-FNNs are employed to construct a trainable network. It seeks to generate codebooks adapting to given CSI and designs codebook-independent decoders free from the exponential increase in signal detection complexity.

We explain the overall system structure, network configuration, and the training and testing processes in sequence. 

\subsection{System Structure of the Proposed Algorithm} \label{S_DL1}

\begin{figure*}[!htpb]
  
  \centering
  \includegraphics[scale=0.95]{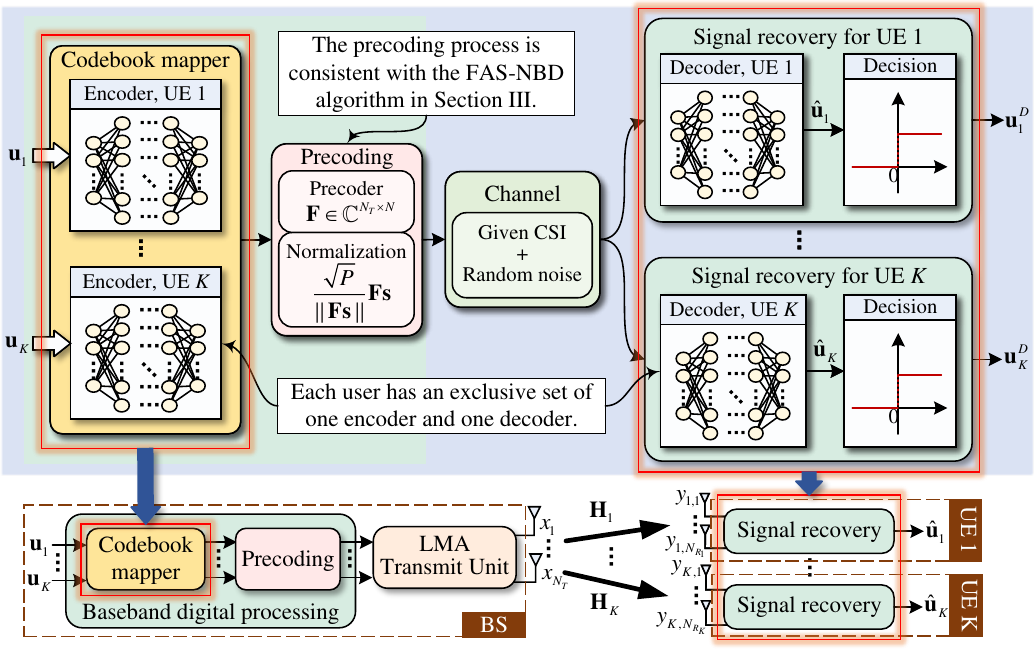}
  \caption{System Structure of the Proposed DL-enhanced Precoding Algorithm (FAS-DL-NBD).}
  \label{fig_DL}
  \end{figure*}

As depicted in Fig.~\ref{fig_DL}, the codebook mapper and signal recovery modules are replaced by FC-FNNs. 
Each user occupies an exclusive encoder that designs user-specific codebooks based on a set of given CSI. It is nested before the precoding module. 
At the receiver, each user necessitates an exclusive decoder to recover signals. 
The encoders and decoders, together with the precoding and channel modules, form the entire neural network. 
The entire network is a regression problem supervised by the recurrence of input information bits, which can be modelled as 
\begin{align}
  \min_\mathbf{\Theta}~\Delta(\mathcal{N}(\mathbf{u};\mathbf{\Theta}),\mathbf{u}), \label{P2}
\end{align}
where $\mathbf{u}=[\mathbf{u}_1^T,\cdots,\mathbf{u}^T_K]^T$ denotes the composite information bit vector for all users, and $\mathbf{\Theta}$ stands for the set of trainable parameters of the network $\mathcal{N}(\cdot)$. $\Delta$ denotes a criterion measuring prediction error and is discussed in detail in Section~\ref{S_DL3}. Notably, elements in the information bit vectors are shifted to be zero-centered, i.e., 0 is represented by $-$0.5 and 1 by 0.5. This is done to avoid the problem of zig-zag paths.

Assume that the encoding and decoding processes of $U_k$ are represented as $\mathbf{E}_k(\cdot)$ and $\mathbf{D}_k(\cdot)$, respectively. 
For $U_k$, the encoded symbol is represented as $\mathbf{s}_k=\mathbf{E}_k(\mathbf{u}_k)$, and the composite symbol vector of all users is represented as $\mathbf{s}=[\mathbf{s}_1^T,\cdots, \mathbf{s}_K^T]^T$. After encoding, the signal is passed into the precoding module which is consistent with the proposed FAS-NBD algorithm. The signal $\mathbf{x}$ to be transmitted is represented as 
\begin{align}
  \mathbf{x} = \frac{\sqrt{P_T}}{\parallel \mathbf{F}\mathbf{E}_k(\mathbf{u}_k) \parallel} \mathbf{F}\mathbf{E}_k(\mathbf{u}_k). \label{DL_S2T}
\end{align}
The signal is then transmitted to the corresponding UE, where decoding and detection are completed. For $U_k$, the prediction vector $\hat{\mathbf{u}}_k$ is expressed as
\begin{align}
  \hat{\mathbf{u}}_k & = \mathbf{D}_k(\mathbf{H}_k\mathbf{x}+\mathbf{n}_k).  \label{DL_SoD}
\end{align} 

As mentioned above, the information bits input by each user are encoded by [$-$0.5, 0.5], so the recovered information bit vector $\mathbf{u}^D_k=[u^D_{k,1}, \cdots, u^D_{k,n_k}]^T$ can be detected with a threshold of 0. For $U_k$, that is
\begin{align}
{u^D_{k,i}} &= \begin{cases}
0,&{\text{if}}\ {u_{k,i}<0} \\ 
{1,}&{\text{otherwise}} 
\end{cases},~\text{for }i=1,2,\cdots,n_k, \label{DL_Detection}
\end{align}
where $u_{k,i}$ denotes the $i$th bit of $\mathbf{u}_k$, and $u^D_{k,i}$ stands for the the $i$th bit of the recovered bits.

\subsection{Configuration Details of the Proposed Network} \label{S_DL3}
An FC-FNN can be treated as a combination of multiple layers of linear and activation functions \cite{FNN}.
In general, each layer can be expressed as
\begin{equation}
  f_l(r_{l-1};\mathbf{\theta}_l)=\zeta (\mathbf{W}_l\mathbf{r}_{l-1}+\mathbf{b}_l), \label{layer}
\end{equation}
where $f_l(\cdot )$ represents the relationship between the input and output of the $l$th layer, $r_{l-1}$ represents the output of the previous layer as the input of the current layer. $\mathbf{W}_l$ and $\mathbf{b}_l$ denote the layer weight and bias, respectively. $\mathbf{\theta}_l=\{\mathbf{W}_l, \mathbf{b}_l\}$ represents the layer parameter set. $\zeta(\cdot)$ denotes the activation function, which is to eliminate the linearity of the network so that the network can better fit a nonlinear model.
The FC-FNN structures of $\mathbf{E}_k(\cdot)$ and $\mathbf{D}_k(\cdot)$ are given in Table~\ref{T1}. Both the encoder and decoder employ fully connected layers as their output layers. The number of hidden layers in an encoder and a decoder is $H_E$ and $H_D$, respectively. $N_h$ denotes the dimension of the corresponding hidden layer. 

\begin{table}[!htpb]
  
  \centering
  \caption{FC-FNN structures and model parameters.}
  \label{T1}
  \begin{tabular}{ccc}
  \hline
  \multicolumn{3}{c}{\textbf{Layout of FC-FNN}}        \\ \hline
  \textbf{Componet} & \multicolumn{1}{|c|}{\textbf{Layer}} & \textbf{Output dimension} \\ \hline
  \multirow{3}{*}{$\text{Encoder}_k$}   & \multicolumn{1}{|c|}{Encoder Input Layer} & $n_k$ \\
  & \multicolumn{1}{|c|}{Hidden Layers} & $N_h$ \\
  & \multicolumn{1}{|c|}{Encoder Output Layer} & $2\times n_k$ \\
  \hline
  \multirow{3}{*}{$\text{Decoder}_k$}   & \multicolumn{1}{|c|}{Decoder Input Layer} & $2\times N_{R_k}$ \\
  & \multicolumn{1}{|c|}{Hidden Layers} & $N_h$ \\
  & \multicolumn{1}{|c|}{Decoder Output Layer} & $n_k$ \\
  \hline
  \end{tabular}
  \end{table}

Notably, since activation functions in an FC-FNN may not support complex-valued numbers, the complex-valued matrices are expressed using block form (i.e., $\mathbf{A} = [\mathbf{A}_r, \mathbf{A}_i]^T$, where $\mathbf{A}_r$ and $\mathbf{A}_i$ represent the real and imaginary parts of $\mathbf{A}$, respectively).
Hence, the dimensions of the transmitter output and receiver input are doubled to preserve the results in block form.  

In this design, for the hidden layers and input layers, the rectified linear unit is chosen as the \emph{activation function}, i.e., $\max (0,x)$.
It is simple and concise, ensuring efficient gradient descent and backpropagation with low computational complexity \cite{ReLU}.
In addition, the \emph{Batch Normalization} is added before the activation layer with the benefits of accelerating the convergence of model training and making the model training process more stable \cite{BN}.

Furthermore, to train a neural network, \emph{Loss functions} are defined to evaluate the prediction performance, and \emph{optimizers} are defined as a guide of backpropagation. 
In this design, the Huber loss function is considered to achieve a robust regression \cite{Huber}:  
\begin{align}
  &L_k = \begin{cases}
  \frac{1}{2}(u_{k,i}-\hat{u}_{k,i})^2,&{\text{if}}\ {|u_{k,i}-\hat{u}_{k,i}|\leq 1} \\ 
   |u_{k,i}-\hat{u}_{k,i}|-\frac{1}{2},&{\text{otherwise}} 
  \end{cases}, \\&~\text{for }i=1,2,\cdots,n_k,\label{Eq_Huber}
  \end{align}
where $L_k$ denotes the loss of $U_k$, and $u_{k,i}$ and $\hat{u}_{k,i}$ stands for the $i$th elements of the information bit vector $\mathbf{u}_k$ and the prediction vector $\mathbf{\hat{u}}_{k}$, respectively. 

Furthermore, we develop a weighted average method to obtain the overall loss for all users.
First, the weights are initialized as a $K$-length row vector with identical values $\frac{1}{K}$.
Then, assuming that the loss vector containing all user losses is $\mathbf{loss}=[loss_1, loss_2, \cdots, loss_K]^T$, the weighted average loss is calculated as 
\begin{align}
  \overline{loss} = \mathbf{w}\cdot \mathbf{loss}. \label{loss_2} 
\end{align}
Further, during the training process, the weight $\hat{\mathbf{w}}$ is updated according to the loss of each user and used for the next epoch. That is
\begin{align}
  \hat{\mathbf{w}} = \frac{\mathbf{loss}^T}{\overline{loss}}, \label{loss_3}
\end{align}
where the sum of the weight vector is constrained to 1, and the network is guaranteed to prioritize users with greater loss values. Here, we choose Adam as the optimizer \cite{Adam}.

\subsection{Training and Testing Processes} \label{S_DL4}

The training and testing process of an FAS-DL-NBD network is summarized as follows.

\begin{algorithm}[!htpb]
  \caption{Training Process.}
  \label{A_train}
  \textbf{Inputs:} $\eta $, $\mathcal{T}_{train}$, $\mathcal{L}$, SNR range, $\mathbf{H}$;\\
  \textbf{Outputs:} {A trained network;}
  \begin{algorithmic}[1]
      \STATE Calculate the BD precoder (see (\ref{BD_matrix})) and initialize network parameters;
      \FOR{$epoch=1$ to $N_{train}$}
      \FOR{$step=1$ to $\frac{N_{train}}{N_{batch}}$}
      \STATE Calculate the signal to be transmitted (see (\ref{DL_S2T})) and add noise with random SNR;
      \STATE Calculate and decode the received signals (see (\ref{DL_SoD}));
      \STATE Update the loss weight (see (\ref{loss_2}) and (\ref{loss_3}));
      \STATE Update the network using Adam with the weighted average loss;
      \ENDFOR
      \ENDFOR
  \end{algorithmic}
  \end{algorithm}

In the training process (\textbf{Algorithm~\ref{A_train}}), bits 0 and 1 are generated randomly with equal probability and are shifted to be zero-centered to form the dataset. The label set $\mathcal{L}$ and the training set $\mathcal{T}_{train}$ are equivalent, which are both comprised of the information bits fed into the system. The network is trained with a given learning rate $\eta$, given CSI, and a specified SNR range. Initially, the BD precoder is calculated, and other network parameters are initialized at random. The network is trained using mini-batches. In each step, we add noise randomly to the received signal to improve the anti-noise capability. The power of added noise is within the given SNR range.
The difference between the sample label and the prediction vector is used for error backpropagation. When the training epoch reaches a specified maximum epoch $N_{train}$, the FAS-DL-NBD network is considered to be well-trained.

\begin{algorithm}[!htpb]
  \caption{Testing Process.}
  \label{A_test}
  \textbf{Inputs:} $\mathcal{T}_{test}$, SNR range, $\mathbf{H}$, the trained model;\\
  \textbf{Outputs:} BER;
  \begin{algorithmic}[1]
      \STATE Calculate BD precoder (see~(\ref{BD_matrix})) and load the trained model;
      \FOR{SNR in SNR range}
      \STATE Calculate the signal to be transmitted (see~(\ref{DL_S2T})) and add noise with given SNR;
      \STATE Calculate and decode the received signals for different UEs (see~(\ref{DL_SoD}));
      \STATE Estimate information bits for different UE based on a given threshold (see~(\ref{DL_Detection}));
      \STATE Count the number of error bits and calculate the BER;
      \ENDFOR
  \end{algorithmic}
  \end{algorithm}

The network testing process (\textbf{Algorithm~\ref{A_test}}) takes the testing set $\mathcal{T}_{test}$, SNR range, $\mathbf{H}$ and the trained FAS-DL-NBD network as inputs, and outputs the BER values with given SNRs. The BER performance is sequentially examined within the specified SNR range. In a slight departure from the training procedure, the noise power of the mmWave channel is determined by the SNR rather than a randomly generated number, and there is no backpropagation of losses. Instead, we obtain the recovered bits by comparing the network outputs to a threshold of 0. 

\vspace{-0.2cm}

\subsection{Algorithm Limitations}

In the design process of the proposed FAS-DL-NBD algorithm, we assume that the CSI is known in advance. The training and testing processes of each FAS-DL-NBD network use the same set of CSI. Although we demonstrate in Section~\ref{S_sim} that the network has a high tolerance for varying ICSI, it is currently limited to quasi-static channels. To handle varying instantaneous channels with large variations, multiple networks must be trained, which could be computationally and time intensive.

To generalize the use case, advanced DL techniques and neural network architectures can be applied \cite{CSI1, CSI2}, or a network dictionary pre-trained on selected channels can be designed.
However, this is not the focus of this paper. In this paper, we particularly focus on discussing the enhancement possibilities of FC-FNNs in terms of codebook design for MU-LMA systems in fading channels.

\section{Discussion on the Advantages of the Proposed FAS-based Algorithms} \label{S_disc1}
Considering the downlink MU-LMA communication system, the proposed FAS-based system offers advantages such as configuration flexibility, the potential to support a large number of users, and algorithm integrity compared with the existing SAS-based system. 

\vspace{-0.2cm}

\subsection{Configuration Flexibility}

To achieve BD precoding, the number of antennas in the SAS-based and FAS-based systems must meet the constraints outlined in Table~\ref{T2}. In comparison to the FAS-based system, the SAS-based system has more stringent restrictions and inflexible system configuration.
\begin{table}[!htpb]
  \setlength{\abovecaptionskip}{3pt}
  \setlength{\belowcaptionskip}{2pt}
  \centering
  \caption{System Dimension Limitations of the FAS-based algorithms and SAS-precoding.}
  \label{T2}
  \begin{tabular}{cc}
  \toprule
  {The FAS-based System} & {The SAS-based System} \\  
  \midrule
    $\left\{\begin{matrix}
      N_R\leq N_T \\ 
      n_k\leq N_{R_k}
    \end{matrix}\right.$ &
    $\left\{\begin{matrix}
      M=\frac{N_T}{K}\in \mathbb{Z}^+ \\ 
      N_{R}\leq M \\
      n_k =  N_{R_k}
    \end{matrix}\right.$\\ 
  \bottomrule
  \end{tabular}
  \end{table}

It indicates that the number of users in an SAS-based system must be divisible by the number of transmit antennas and is further limited by the total number of receive antennas. An FAS-based system, on the other hand, diminishes the limitations imposed by an SAS-based system, and in turn, supports varying numbers of users dynamically.

\subsection{Potential to Support a Large Number of Users}

Given the system constraints outlined in Table~\ref{T2}, the FAS transmitter assists in facilitating a large number of users. 

\subsubsection{Fewer transmit antennas are required}
For simplicity, assume the number of receive antennas $N_{R_k}$ for each user is equal. Given the constraints in Table~\ref{T2}, the number of users $K$ for the FAS-based system is constrained by
\begin{align}
  K = \frac{N_R}{N_{R_k}} \leq \left\lfloor \frac{N_T}{N_{R_k}} \right\rfloor. \label{FAS-K}
\end{align}
Meanwhile, the user number of the SAS-based system is constrained by
\begin{align}
  K &= \frac{N_R}{N_{R_k}} \leq \frac{M}{N_{R_k}} = \frac{N_T}{K\cdot N_{R_k}}.
\end{align}
As a result, 
\begin{align}
  K &\leq \left\lfloor \sqrt{\frac{N_T}{N_{R_k}}} \right\rfloor. \label{SAS-K}
\end{align}
The configuration is further constrained by $\frac{N_T}{K}\in \mathbb{Z}^+$.

\begin{table*}[!htpb]
  
  \centering
  \caption{The Number of Transmit Antennas per User with Different Numbers of Transmit Antennas and Users ($N_{R_k}=2$).}
  \label{TDoF}
  \begin{tabular}{c|cccccccccc}
    \Xhline{1.2pt}
    \multirow{2}{*}{\diagbox{$N_T$}{$K$}} & \textbf{2} & \textbf{3} & \textbf{4} & \textbf{5} & \textbf{6} & \textbf{7} & $\cdots$ & \textbf{12} & $\cdots$ & \textbf{144} \\
    \cline{2-11}
    & \multicolumn{10}{c}{\textbf{FAS-based System } \textbf{\textbar} \textbf{ SAS-based System}}          \\
  \hline
  \textbf{24} & 24 \textbar~12 & 24 \textbar~8 &  24 \textbar~$\times$ &  24 \textbar~$\times$ & 24 \textbar~$\times$  &  24 \textbar~$\times$ & $\cdots$ & 24 \textbar~$\times$ & $\cdots$ & $\times$ \textbar~$\times$   \\
  \textbf{288} & 288 \textbar~144 & 288 \textbar~96 & 288 \textbar~72  & 288 \textbar~$\times$  & 288 \textbar~48  &  288 \textbar~$\times$ & $\cdots$ & 288 \textbar~24 & $\cdots$ & 288 \textbar~$\times$ \\
  \Xhline{1.2pt}
  \end{tabular}
  \end{table*}

Referring to (\ref{FAS-K}) and (\ref{SAS-K}), the FAS-based system supports a greater number of users than the SAS-based system, and as the number of users increases, this difference will be significant. Assume $N_{R_k}=2$. Table~\ref{TDoF} lists the number of transmit antennas per user, based on various combinations of transmission antenna numbers $N_T$ and user number $K$. The values for the FAS-based system are listed on the left side of each cell, while those for the SAS-based system are on the right. Notably, a ``$\times$'' denotes that the system cannot support the given combination of $N_T$ and $K$. It can be seen that an FAS-based system with $N_T=24$ can support up to 12 users. However, an SAS-based system with the same configuration can only support 3 users. To support the same number of 12 users in an SAS-based system, at least 288 transmit antennas are required.

\subsubsection{Greater Gain in Degrees of Freedom}
    
The FAS-based system has a greater degree-of-freedom gain in comparison to the SAS-based system. As users in the FAS-based system share the entire antenna array and transmit signals independently, the number of available transmit antennas per user for the FAS-based system is the number of transmit antennas, that is
\begin{align}
  M=N_T.
\end{align}
Unlike the FAS-based systems, users in the SAS-based system occupy only a portion of the antenna array, resulting in each user being assigned $\frac{N_T}{K}$ transmit antennas.
Thus, the available range for the SAS-based system is bounded by
\begin{align}
  \left\lceil \sqrt{N_{R_k}N_T} \right\rceil \leq M \leq N_T ,
\end{align}
where the lower limit of the inequality denotes the minimum number of transmit antennas available to each user in a fully loaded system (i.e., the number of users in the system reaches the maximum), while the upper limit corresponds to a system with only 1 user.

Assume $N_{R_k}=2$. Table~\ref{TDoF} lists the number of transmit antennas per user of FAS-based and SAS-based systems for varying numbers of users when $N_T=24$ and 288, respectively. 
The difference between the FAS-based and SAS-based systems in the number of antennas per user increases as the number of users rises. The FAS-based system's degree-of-freedom gain will result in performance improvements \cite{book2}, and its advantages over the SAS-based system will become apparent as its user base expands. This is demonstrated in Section~\ref{S_sim}. 

\subsubsection{Robustness with Varying Numbers of Users}

Apart from affecting the number of transmit antennas per user, the increasing number of users also affects the MUI cancellation process. Both the proposed FAS-based and the SAS-precoding algorithms consider constructing precoders in the null space of non-target user channels. However, as the number of users increases, the dimension of the null space (i.e., the rank of the matrix used to achieve MUI cancellation) decreases. This results in a decrease in received signal power when the intended user's channel is projected on the null space, consequently leading to degraded BER performance. Referring to (\ref{Eq_span}), the null space dimension of $U_k$ (i.e., the rank of $\widetilde{\mathbf{V}}_k^0$) of the FAS-based algorithm is represented as
\begin{align}
  r_{\text{FAS-NBD}}^k=N_T-\sum_{i\neq k}^{K} N_{R_k}.
\end{align}
Similarly, the null space dimension of $U_k$ of the SAS-precoding algorithm in \cite{ref_n8} is represented as
\begin{align}
  r_{\text{SAS-precoding}}^k=\frac{N_T}{K}-\sum_{i\neq k}^{K} N_{R_k}. \label{SAS-rank} 
\end{align}
Referring to (\ref{SAS-rank}), the MUI cancellation process of the SAS-precoding is prone to changes in the number of users $K$. In contrast, owing to the constant degrees of freedom, the proposed FAS-based algorithm is more stable as the number of users changes.

\subsection{Algorithm Integrity: Joint Design of Precoder and Combiner, Signal Energy Maximization}
The FAS-based algorithm compensates for the lack of combiner design and the unstable algorithm performance in the SAS-precoding algorithm.

\subsubsection{Joint Design of Precoder and Combiner}
The SAS-precoding does not include the design of combiners, so the system is restricted to situations where the number of transmitted bits and the number of receive antennas are equal. The proposed FAS-NBD relaxes system constraints by designing the precoder and combiner jointly.

\subsubsection{Signal Energy Maximization}
Assume the precoding matrix of $U_k$ in the SAS-precoding is $\mathbf{T}_k$. It can be expressed as 
\begin{align}
  \mathbf{T}_k = \mathbf{V}_k\mathbf{B}_k, 
\end{align}
where $\mathbf{V}_k$ block diagonolizes the MUI, and $\mathbf{B}_k$ is a semi-unitary matrix with full column rank satisfying $\mathbf{B}_k^H\mathbf{B}_k=\mathbf{I}$. It adapts the dimension of $\mathbf{V}_k$ to transmit the desired number of information bits.
However, without a given criterion, this matrix is randomly generated, and it is likely to rotate the encoded signal orthogonal to the channel that 
\begin{align}
  \mathbf{H}_k\mathbf{T}_k = \mathbf{H}_k\mathbf{V}_k\mathbf{B}_k=\mathbf{0}. 
\end{align}
Under such a situation, the BER performance will seriously deteriorate. 
In contrast, the proposed FAS-NBD ensures algorithm robustness and enhances performance by maximizing signal energy in the direction corresponding to the equivalent channel (as shown in (\ref{Eq_span})). 
Moreover, with the improved signal direction, FAS-based systems tend to have a larger pairwise distance between constellation points at the receiver (i.e., $\parallel \mathbf{H}_k\mathbf{F}_k\mathbf{s}_i - \mathbf{H}_k\mathbf{F}_k\mathbf{s}_j\parallel, \forall \mathbf{s}_i, \mathbf{s}_j \in {\mathcal{S}_{n_k, P}}, i\neq j$), resulting in improved robustness against disturbances.

\section{Discussion on the Convergence Superiority of the Proposed DL-enhanced Algorithm} \label{S_disc2}

The concept of end-to-end learning allows for the utilization of an E2E-based (i.e., FAS-E2E) framework to simultaneously address the MUI cancellation and codebook-related problems in the downlink MU-LMA system.
However, this multi-task framework sparks convergence difficulty, resulting in a waste of training resources. 
In this section, we compare the FAS-E2E framework with the proposed FAS-DL-NBD framework. We demonstrate the auxiliary effect of NBD on network training and the advantages of the joint algorithm compared to the FAS-E2E algorithm. 

\begin{table}[!htpb]
  
  \centering
  \caption{The configuration of an E2E-transmitter.}
  \label{T_DL_P}
  \begin{tabular}{ccc}
  \hline
  \multicolumn{3}{c}{\textbf{Layout of FC-FNN}}        \\ \hline
  \textbf{Componet} & \multicolumn{1}{|c|}{\textbf{Layer}} & \textbf{Output Dimension} \\ \hline
  \multirow{3}{*}{E2E transmitter}   & \multicolumn{1}{|c|}{Input Layer} & $2\times N$ \\
  & \multicolumn{1}{|c|}{Hidden Layers} & $N_h$ \\
  & \multicolumn{1}{|c|}{Output Layer} & $2\times N_T$ \\
  \hline
  \end{tabular}
  \end{table}

The FAS-E2E can be achieved by replacing the FAS-DL-NBD transmitter (including the encoders and the precoding module) with a single FC-FNN. The configuration of this \emph{E2E-transmitter} is shown in Table~\ref{T_DL_P}. Other components, such as the normalization, the channel model, and the decoders for each UE are consistent with the configuration in the proposed FAS-DL-NBD algorithm (Table~\ref{T2}). 
It receives the information bits from all users and forwards them to be normalized directly. 

\begin{figure}[!htpb]
  
  \centering
  \includegraphics[scale=0.8]{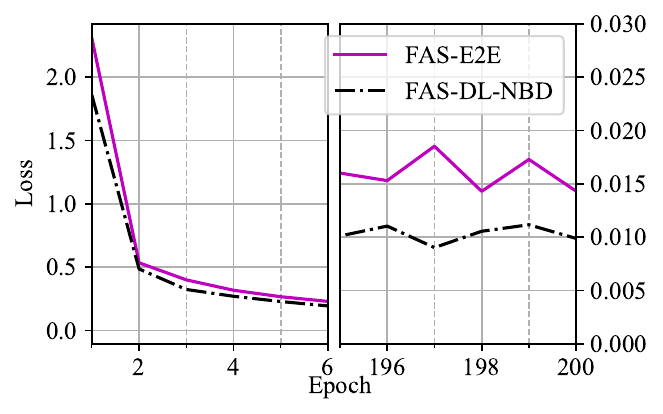}
  \caption{Comparison of loss as a function of training epoch between FAS-DL-NBD and FAS-E2E, where $K=2$, $N_T=18$, $n_k=2$, $N_{R_k}=2$ and $N_{ray}=3$.}
  \label{fig_loss}
  \end{figure}

Here, we consider a system with $K=2$, $N_T=18$, $n_k=2$, $N_{R_k}=2$ and $N_{ray}=3$. Two networks based on these two frameworks (i.e., FAS-E2E and FAS-DL-NBD) are trained with the same parameters set (i.e., Set I in Table~\ref{TP}). Fig.~\ref{fig_loss} compares their training losses. 
It demonstrates that the FAS-DL-NBD network has a smaller initial loss and converges to 0.010 faster around the epoch of 200. This is the result of the supplementary effect of the computed BD precoder on MUI cancellation.
In contrast, the loss of the FAS-E2E network plateaus at around 0.015. The network fails to converge to a lower value because it is struggling to suppress MUI and optimize prediction accuracy at the same time.
Comparisons of their BER performance are presented and illustrated in Section \ref{S_sim}.

\section{Simulation Results} \label{S_sim}

All algorithms are implemented on an Intel i7-1165G7 CPU with 16 GB RAM using Python. One NVIDIA GeForce MX450 GPU is used to train the neural networks. 
Table~\ref{TP} describes the training parameters for the models used in the simulation. Each column represents a parameter set. 
Moreover, in the subsequent simulation process, both the FAS-NBD and the SAS-precoding algorithms employ the conventional LMA codebooks (i.e., codewords uniformly distributed on a multidimensional hypersphere using K-means clustering \cite{KMeans}). On the other hand, the per-user codebook of the FAS-DL-NBD algorithm is the training result of the corresponding encoder $\mathbf{E}_k(\cdot)$.
Furthermore, we assume that the system employs a training-based channel estimation method with minimum mean-square error at each UE to obtain channel information \cite{CE1}, which is then transmitted via error-free uplink channels to the BS.

\begin{table}[!htpb]
  
  \setlength{\tabcolsep}{4pt}
  \centering
  \caption{Training Parameters}
  \label{TP}
  {\begin{tabular}{ccccc}
  \toprule
  {\textbf{Parameter}} & {\textbf{Set I}} & {\textbf{Set II}} & {\textbf{Set III}} & {\textbf{Set IV}} \\  
  \midrule
  {Number of user, $K$} & {2} & {3} & {4} & {2} \\ 
  \midrule
  {Dimension of hidden layer, $N_h$} & {128} & {128} & {128} & {128} \\ 
  \midrule
  {Number of hidden layer for $\mathbf{E}_k$, $H_E$} & {3} & {3} & {3} & {3} \\ 
  \midrule
  {Number of hidden layer for $\mathbf{D}_k$, $H_D$} & {2} & {2} & {2} & {2} \\ 
  \midrule
  {Batch Size} & {100} & {100}  & {100} & {100} \\ 
  \midrule
  {Number of Samples, $|\mathcal{T}_{train}|$} & {$10^3$} & {$10^4$} & {$10^4$} & {$10^{4}$} \\ 
  \midrule
  {Number of training epochs, $N_{train}$} & {200} & {300} & {300} & {400} \\ 
  \midrule
  {Learning Rate, $\eta$} & {$10^{-3}$} & {$10^{-3}$} & {$10^{-3}$} & {$10^{-3}$} \\   
  \midrule
  {SNR Range} & \multicolumn{4}{c}{0 to 15dB} \\ 
  \bottomrule 
  \end{tabular}}
  \end{table}

\begin{figure}[!htpb]	
	\centering
	\includegraphics[scale=0.85]{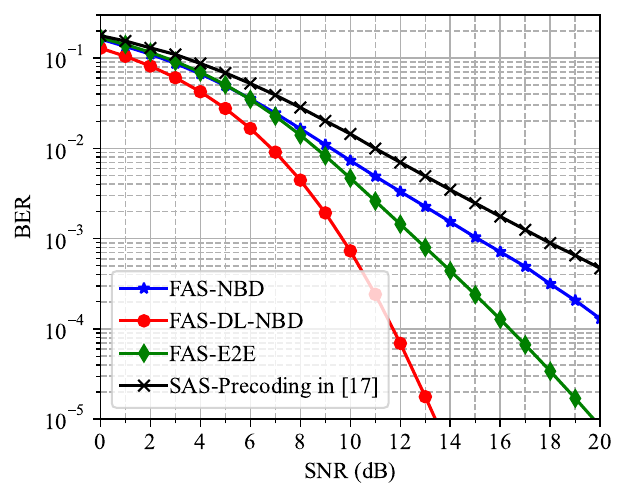}
	\caption{Comparison of the BER performance in the 3-user downlink MIMO LMA system employing FAS-NBD, FAS-DL-NBD, SAS-precoding and FAS-E2E where $K=3$, $N_T=24$, $n_k=N_{R_k}=2$ and $N_{ray}=3$.}\label{fig: BER}
\end{figure}

Assuming full CSI is known by the system, Fig.~\ref{fig: BER} depicts the BER performance of the proposed two algorithms (i.e., FAS-NBD and FAS-DL-NBD). They are compared with the existing SAS-precoding in \cite{ref_n8} and the one-step FAS-E2E described in Section~\ref{S_disc2}. We consider a system with $K=3,~N_T=24,~N_{ray}=3,~\text{and}~N_{R_k}=n_k=2$. Both the FAS-DL-NBD network and the FAS-E2E network are trained with the parameters in Set II (Table~\ref{TP}). 
The algorithms were tested using varying instantaneous channels, and the FAS-DL-NBD and FAS-E2E networks were trained for each channel. 
For the fairness of the comparison, the transmit power for all of the algorithms is set to be the same and $P_T=N_T$.

The proposed FAS-NBD algorithm outperforms the existing SAS-precoding algorithm in terms of BER performance. It achieves 3 dB gain at $\text{BER}=10^{-3}$. This performance improvement is from the degree-of-freedom gain and the maximized signal energy.
The FAS-DL-NBD algorithm, meanwhile, exhibits the best BER performance. It is a successive enhancement to FAS-NBD, and the BER performance at $10^{-4}$ is further enhanced by 5 dB. By training independent codebooks for each user, it provides codewords that are robust to the channel effects. 
Furthermore, the inferior BER performance of the FAS-E2E algorithm in comparison to FAS-DL-NBD serves as evidence of the negative impact caused by the convergence problem.

\begin{figure}[h]
  
  \centering
  \includegraphics[width=\linewidth]{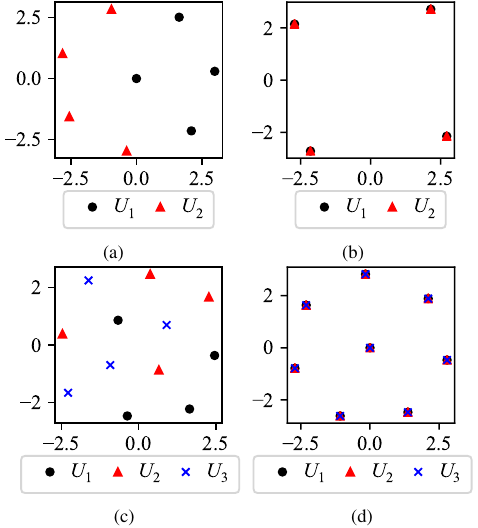}
  \caption{Constellation diagrams of FAS-DL-NBD and conventional algorithms. (a) FAS-DL-NBD constellation for 2 users; (b) Conventional constellation for 2 users; (c) FAS-DL-NBD constellation for 3 users; (d) Conventional constellation for 3 users.}
  \label{Cons}
\end{figure}

Fig.~\ref{Cons}a and Fig.~\ref{Cons}c show the constellation diagrams of FAS-DL-NBD systems with 2 and 3 users, respectively. Each FAS-NBD constellation presented corresponds to a given set of CSI. As seen from the figure, unlike the conventional codebooks (Fig.~\ref{Cons}b and \ref{Cons}d) used in FAS-NBD and SAS-precoding algorithms, the FAS-DL-NBD algorithm designs individual codebooks for users which are robust to channel effects. 

In addition, to illustrate the advantages of the FAS-based algorithms with increasing user numbers, we consider 3 multiuser systems. The number of users is 2, 3, and 4, respectively. Other parameters are same that $N_T=36,~N_{ray}=3,\text{ and}~n_k=N_{R_k}=2$. 
The corresponding FAS-DL-NBD networks are trained with the parameters in Set I, II, and III (Table~\ref{TP}), respectively. 
In the SAS-based system, the number of transmit antennas available to each user decreases from 18 to 9 as the number of users increases, whereas in the FAS-NBD system, this number is fixed at 36, as users achieve independent transmission on a shared antenna array.

\begin{figure}[!htpb]
  
  \centering 
  \includegraphics[scale=0.85]{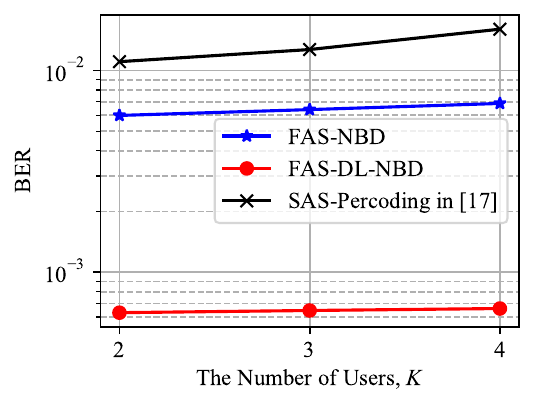}
  \caption{BER performance comparison as a function of user number $K$ between FAS-NBD and FAS-DL-NBD, where $N_T=36$, $n_k=N_{R_k}=2$ and $N_{ray}=3$.}
  \label{fig_K}
  \end{figure}
  
Fig.~\ref{fig_K} shows the BER performance with the increase of user number $K$ at $\text{SNR}=10~\text{dB}$. The performance was tested with varying instantaneous channels, and the networks were trained for each channel. As the number of users increases, the performance advantage of the proposed algorithms increases compared to the SAS-precoding algorithm. 
The performance degradation of SAS-precoding is attributed to the reduced degrees of freedom per user. On the other hand, the FAS-based algorithms consistently exhibit superior performance owing to constant degrees of freedom and signal transmission with energy maximized. This result confirms the robustness of the FAS-based algorithms under varying numbers of users. Furthermore, the FAS-DL-NBD algorithm consistently outperforms its counterparts due to its channel-adaptive codewords.

\begin{figure}[!htpb]
  
  \centering
  \subcaptionbox{\label{fig:1}}{\includegraphics[scale=0.85]{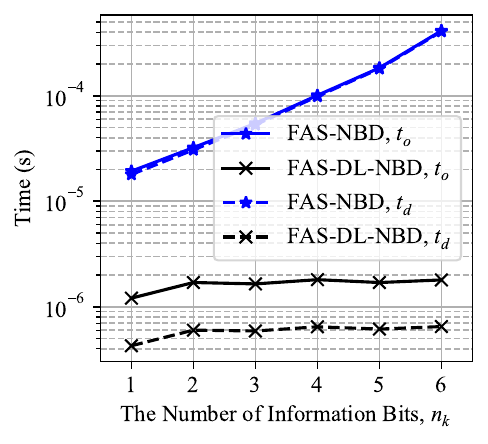}}
  \subcaptionbox{\label{fig:2}}{\includegraphics[scale=0.85]{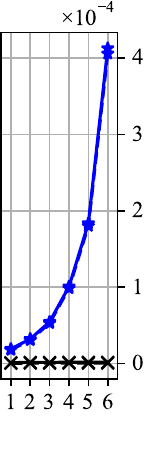}}
  \caption{Comparison of the overall communication time $t_o$ and the signal detection time $t_d$ as a function of the number of information bits between FAS-NBD and FAS-DL-NBD, where $K=2$, $N_T=16$, $N_{R_k}=6$ and $N_{ray}=3$. (a) Time axis in log-scale; (b) Time axis in regular-scale.}
  \label{fig_time}
\end{figure}

In addition, Table~\ref{CC} summarizes the computational complexity of FAS-NBD and FAS-DL-NBD for $U_k$. The detection complexity of FAS-NBD and FAS-DL-NBD can be simplified as proportional to $O\left(2^{n_k}\times n_k\right)$ and $O\left(n_k\right)$, respectively when the structures of the MU-LMA system and the FAS-DL-NBD neural network are fixed (i.e., $N_{R_k}$, $N_T$, $N_h$, $H_D$ and $H_E$ are constant).
\begin{table*}[!htpb]
  
  \centering
  \caption{Computational Complexity of FAS-NBD and FAS-DL-NBD for $U_k$}
  \label{CC}
  \begin{tabular}{c}
  \toprule
  {\textbf{Computational Complexity of FAS-NBD for $U_k$}} \\
  \midrule
  {$O\left(\underbrace{\overbrace{2^{n_k}(N_T\times n_k+N_{R_k}\times N_T+2N_{R_k})}^{\text{Detection complexity}}+n_k\times N_T}_{\text{Overall communication complexity}}\right)$} \\ 
  \midrule
  {\textbf{Computational Complexity of FAS-DL-NBD for $U_k$}} \\
  \midrule
  {$O\left(\underbrace{\overbrace{n_k(N_h+2)+(H_D-1)(N_h^2+2N_h)+(2N_{R_k}+2)N_h}^{\text{Detection complexity}}+2N_h+(H_E-1)(N_h^2+N_h)+n_k(2N_h+1+N_T)}_{\text{Overall communication complexity}}\right)$} \\ 
  \bottomrule
  \end{tabular}
  \end{table*}
Besides, the prediction process in FAS-DL-NBD can be accelerated by parallelism, allowing its computational complexity to be further compressed in implementation. Fig.~\ref{fig_time} intuitively illustrates the trends of the overall communication time (denoted as $t_o$) and the detection time (denoted as $t_d$) of the two algorithms at 10 dB as a function of the number of transmitted bits $n_k$.
Other parameters are same that $K=2,~N_T=16,~N_{ray}=3,\text{ and}~N_{R_k}=6$. The FAS-DL-NBD networks are trained with parameters in Set IV (Table~\ref{TP}).

The graph represents the average time per user for the corresponding process. 
The logarithmic time-axis in Fig.~\ref{fig:1} provides greater clarity, while the regularly scaled time-axis in Fig.~\ref{fig:2} provides intuitive trends.
Due to the reduced detection complexity and the linear relationship with the number of information bits, the time consumption of the trained FAS-DL-NBD network is significantly lower than that of the FAS-NBD system and remains stable as the number of information bits increases.
In contrast, the time consumption of the FAS-NBD system increases exponentially as the number of information bits rises. 
At $n_k=6$, the FAS-DL-NBD algorithm exhibits a performance improvement of $230\times$ in overall communication time and a $625\times$ improvement in detection time over the FAS-NBD algorithm. 

\begin{figure*}[!htpb]
  
  \centering
  \subcaptionbox{\label{fig:CSI_DL}}{\includegraphics[scale=0.85]{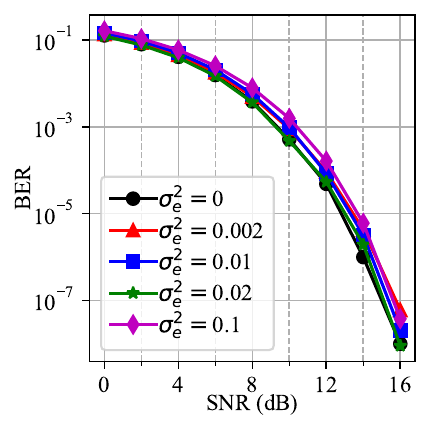}}
  \subcaptionbox{\label{fig:CSI_BD}}{\includegraphics[scale=0.85]{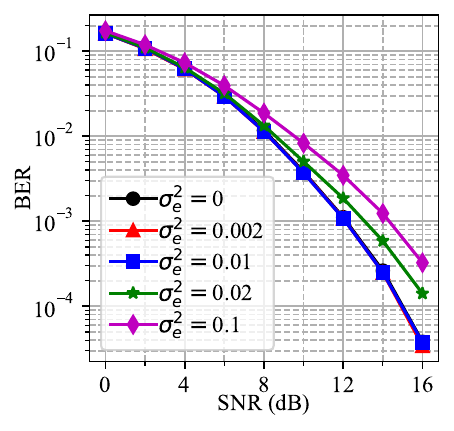}}
  \subcaptionbox{\label{fig:CSI_FX}}{\includegraphics[scale=0.85]{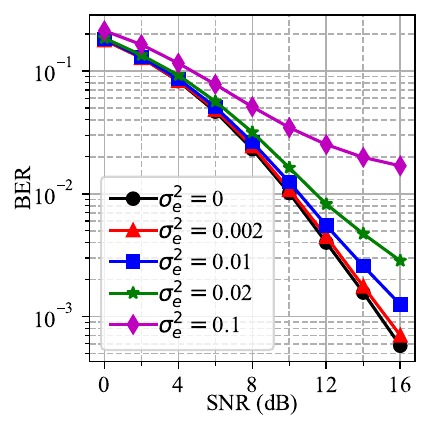}}
  \caption{Comparison of the BER performance with imperfect CSI, where $K=2$, $N_T=12$, $n_k=N_{R_k}=2$ and $N_{ray}=3$. (a) FAS-DL-NBD; (b) FAS-NBD; (c) SAS-precoding in \cite{ref_n8}.}
  \label{CSI}
\end{figure*}

Furthermore, the performance of the three algorithms is compared to that of ICSI. The channel estimation error is modelled as $\mathcal{CN}(0,\sigma^2_e)$ {\cite{CE1, CE2}}. 
It may lead to residual MUI and render the FAS-DL-NBD algorithm incapable of training optimal codebooks.
As shown in Fig.~\ref{CSI}, the performance of the SAS-precoding algorithm is seriously damaged by ICSI due to the randomness of the precoder design. 
In contrast, the BER of the FAS-NBD algorithm varies within a small range and this confirms the robustness of the algorithm to disturbances. Such robustness is attributed to the degree-of-freedom gain and signal energy maximization.
Nonetheless, the FAS-DL-NBD algorithm utilizes estimated CSI to train an approximately optimal constellation, ensuring robustness even with a certain degree of channel estimation error.

\section{Conclusions} \label{S_con} 

Communication systems employing LMAs alleviate the increasing system complexity and RF chain cost suffered by MIMO systems. 
In this paper, we have developed a new system framework employing an FAS transmitter of LMA for MU mmWave downlink transmission. 
The proposed FAS-based MU-LMA system addresses the structure-related problems in the existing SAS-based systems with increased degree-of-freedom gains and increased configuration flexibility. 
Apart from that, the FAS-based system breaks the maximum number of users that can be supported and achieves advantages in systems with varying numbers of users. 
Accordingly, we have proposed two algorithms (i.e., FAS-NBD and FAS-DL-NBD) to address the precoding and the codebook-related problems in turn. 

The proposed FAS-NBD algorithm is an optimization based on conventional BD. 
In addition to eliminating MUI in the downlink scenario, the FAS-NBD algorithm adapts to the LMA system structure with a constant power constraint and thus ensures power efficiency. 
Moreover, it implements a one-step design of a set of precoders and combiners, thereby resolving the SAS-precoding algorithm's combiner shortage. 
We have shown that it gives rise to a better BER performance than the existing SAS-precoding algorithm and is more robust when the CSI estimation is imperfect. 

We have also observed performance degradation due to forced normalization. 
Furthermore, the signal codebook of the FAS-NBD is inflexible and the ML detection complexity increases exponentially with the increase of the number of information bits. 
These problems are well solved in the proposed FAS-DL-NBD algorithm. 
By nesting FC-FNNs at the transmitter and receivers respectively, the FAS-DL-NBD network seeks to generate codebooks robust to fading channels as well as achieves low-complex signal detection independent of the codebook size. 
In this way, the proposed FAS-DL-NBD provides a codebook design method with high bit-level precision and stimulates the possibility of transmitting a large number of information bits at a time. 
Furthermore, we also show that in contrast to the conventional one-step end-to-end network, the FAS-DL-NBD network holds promising superiority in network convergence. 

Within the FAS-DL-NBD network design, dedicated training is carried out for each unique CSI. 
It is of interest to study a similar scenario but adapt the neural network to varying instantaneous channels. 
One approach is to leverage more advanced DL techniques and innovative network architectures to improve the model's flexibility and ability to generalize across different channel scenarios. Alternatively, it may be possible to pre-train a network dictionary on selected channels to improve online prediction efficiency.
Last but not least, the channel capacity of the MU-LMA systems is in general an open problem, and in this context, the capacity of the system with constant power constraints is of particular interest.


 
%
\bibliographystyle{IEEEtran}


 




\vfill

\end{document}